\documentclass[twoside,10pt]{article}
\usepackage{amsmath,amssymb}
\usepackage{xspace}
\usepackage{tam13}

\usepackage{lineno}

\title{The Pierre Auger Observatory:  \\ \indent
results on the highest energy particles}
\shorttitle{The Pierre Auger Observatory}
\authors{R.~Concei\c{c}\~{a}o$^1$\email{ruben@lip.pt},
{\it for the} Pierre Auger Collaboration$^{2}$}
\shortauthors{R.~Concei\c{c}\~{a}o, {\it for the} Pierre Auger Collaboration}
\affiliations{
$^1$ LIP, Av. Elias Garcia, 14-1, 1000-149 Lisbon, Portugal
\\
$^2$ Observatorio Pierre Auger, Av. San Mart\'in Norte 304,\\
 (5613) Malarg\"{u}e, Mendoza, Argentina \\
{\small (Full author list: http://www.auger.org/archive/author\_2013\_03.html)}
}

\abstract{
The Pierre Auger Observatory has been designed to investigate the most energetic particles known, the ultra high energy cosmic rays.
The observatory, covering an area of 3000 km$^2$, combines two different detection techniques to study the huge particle showers created by the interaction of primary cosmic rays with the atmosphere.
The analysis of the showers allows one to extract information on the nature of the primary cosmic rays, as well as their origin. Moreover, the study of the interaction of these particles with the atmosphere offers a unique window to study particle physics at an energy more than one order of magnitude above the current highest energy human-made accelerator.
In this contribution selected results are presented, with a focus on the primary mass composition, the determination of the number of muons, which is sensitive to the shower hadronic interactions, and the measurement of the proton-air cross-section at $\sqrt{s} = 57$ TeV. For the last topic, a link with the proton-proton cross-section measurements using accelerators will be made.
Results on the cosmic ray energy spectrum and on searches for ultra high energy photons and neutrinos, will also be discussed.
}

\begin{document}

\maketitle

\section{Introduction}
Ultra High Energy Cosmic Rays (UHECRs) are the most energetic particles known in the Universe. Although they were first reported 50 years ago~\cite{Linsley}, their origin and nature are still open questions. The main reason resides on the fact that direct detection is not possible, mainly due to the very low fluxes. However, as these particles reach the Earth, they interact with the nuclei of the atmosphere, producing huge avalanches of particles, the Extensive Air Showers (EAS). These cascades can be detected either using the secondary charged particles that reach the ground or, on moonless nights, one can sample the longitudinal shower profile through the fluorescence light emitted by the nitrogen molecules that were excited by the shower propagation in the atmosphere. 

UHECRs are a very interesting scientific subject, not only as they possibly probe some of the most energetic processes in the Universe, but also because their interaction with the atmosphere can occur at energies up to $\sqrt{s} \sim 400$ TeV, well beyond any current man-made accelerator. Therefore, they can also be used to investigate hadronic interactions at the highest energies.

\section{The Pierre Auger Observatory}
The Pierre Auger Observatory~\cite{PAO} was conceived to study UHECRs. It was built in Pampa Amarilla, Mendoza, Argentina, and combines two independent techniques to detect EAS: a Surface Detector array (SD) that samples charged particles arriving at the ground, and Fluorescence Detectors (FD) that record the fluorescence light. 

The SD is composed by 1660 water Cherenkov stations, placed in a triangular grid of $1.5$ km spacing, covering a total of 3000 km$^2$. Each station contains 12 tons of water, and is equipped with 3 PMTs that collect the Cherenkov light emitted by charged particles as they cross the water volume. A solar panel, GPS timing and radio communications make the SD stations fully autonomous. 

\begin{figure}[h]
\centering
\includegraphics[width=0.55\textwidth]{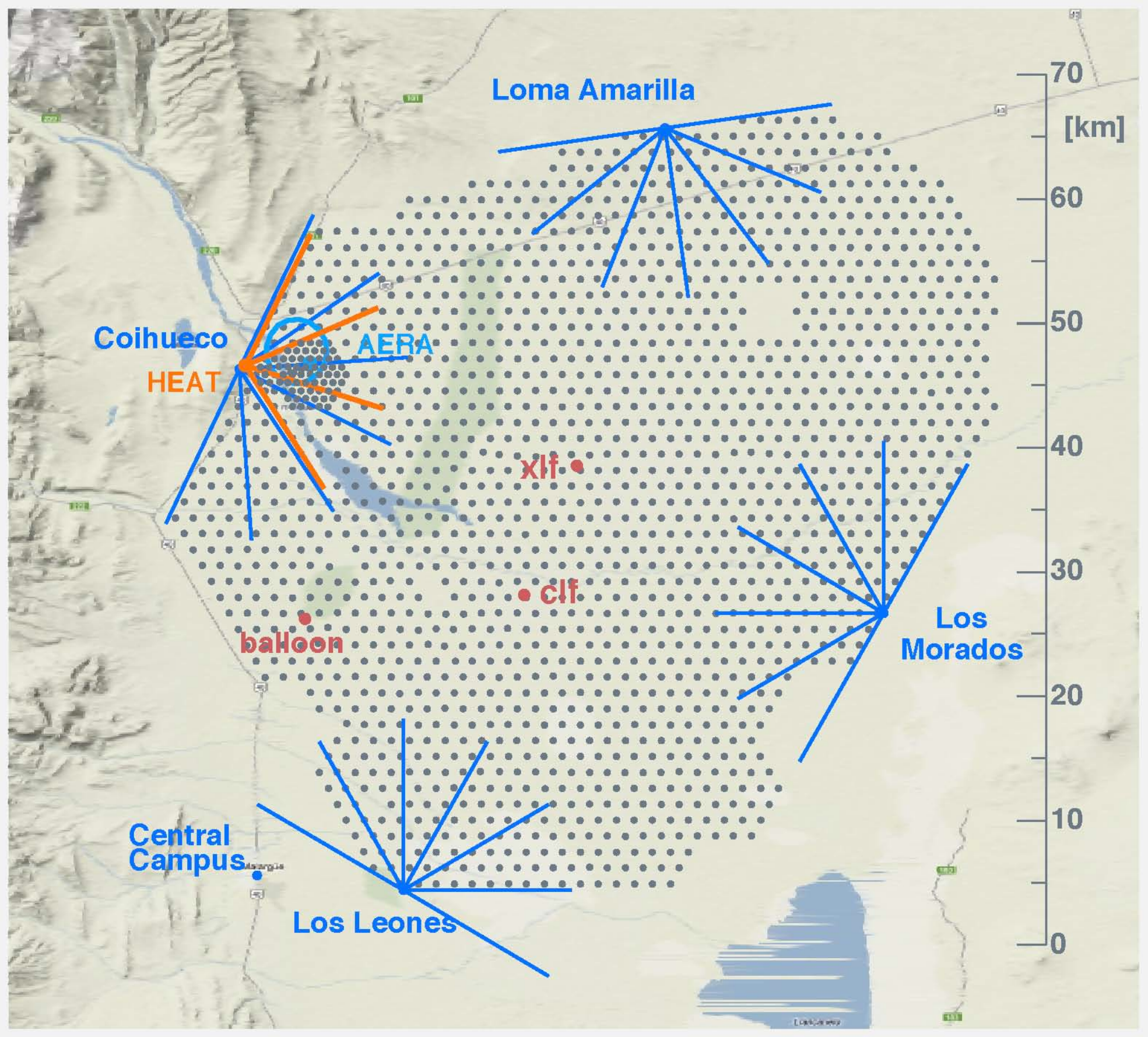}
\caption{The Pierre Auger Observatory located near Malarg\"{u}e, Province of Mendoza, Argentina. The FoV of the fluorescence telescopes (blue/orange radial lines) covers the water Cherenkov surface detector array (dots).
\label{f:meth}}
\end{figure}

The direction of the primary particle is extracted from the footprint on ground, in particular, through the times of arrival of the shower front at the ground.The current timing allows one to reconstruct the shower direction to better than $1^\circ$ at the highest energies. As for the energy of the shower, it is determined using the estimated signal at $1000$ m from the shower core. The SD array~\cite{SD} becomes fully efficient at energies above $E = 3 \times 10^{18}$ eV, both for proton and iron induced showers with zenith angle $\theta < 60^\circ$.

The SD array is overlooked by $27$ Fluorescence Telescopes distributed over 5 sites~\cite{FD}. Each telescope has a $30^\circ \times 30^\circ$ field of view (FoV), and is composed of a UV filter, and a spherical mirror that reflects the light into a camera with $440$ PMTs where it is detected. The integration of the fluorescence light profile allows one to obtain a nearly calorimetric measurement of the shower energy.

While the SD operates continuously, the FD only collects data on moonless nights, resulting in a duty cycle of around $15\%$. The simultaneous operation of the SD and the FD is one of the most important features of the Pierre Auger Observatory. It allows one to improve the geometrical reconstruction of the shower and, even more importantly, to calibrate the energy estimator of the SD, which is model dependent, with the energy reconstructed by the FD. The calibration is performed using high quality hybrid events\footnote{events that can be reconstructed independently with the SD and the FD}. At present $839$ high quality events are considered. The SD energy estimator, $S(1000)$, the signal at $1000$ m from the shower core, is obtained by fitting the lateral distribution function (LDF) of the shower on ground with a modified NKG function. Shower attenuation depends on the zenith angle of the primary particle. So, to account for this a constant intensity method is used, and $S(1000)$ is turned into the equivalent signal at $\theta = 38^\circ$. Finally, the calibration plot between the SD energy estimator and the energy measured by the FD is built. The main systematic uncertainty on the energy measurement comes from the FD energy scale and amounts to around $22\%$~\cite{calib}.

The hybrid event analyses provide the observatory with an additional handle on the understanding of EAS physics.

\section{Energy Spectrum}
\begin{figure}[h]
\centering
\includegraphics[width=0.85\textwidth]{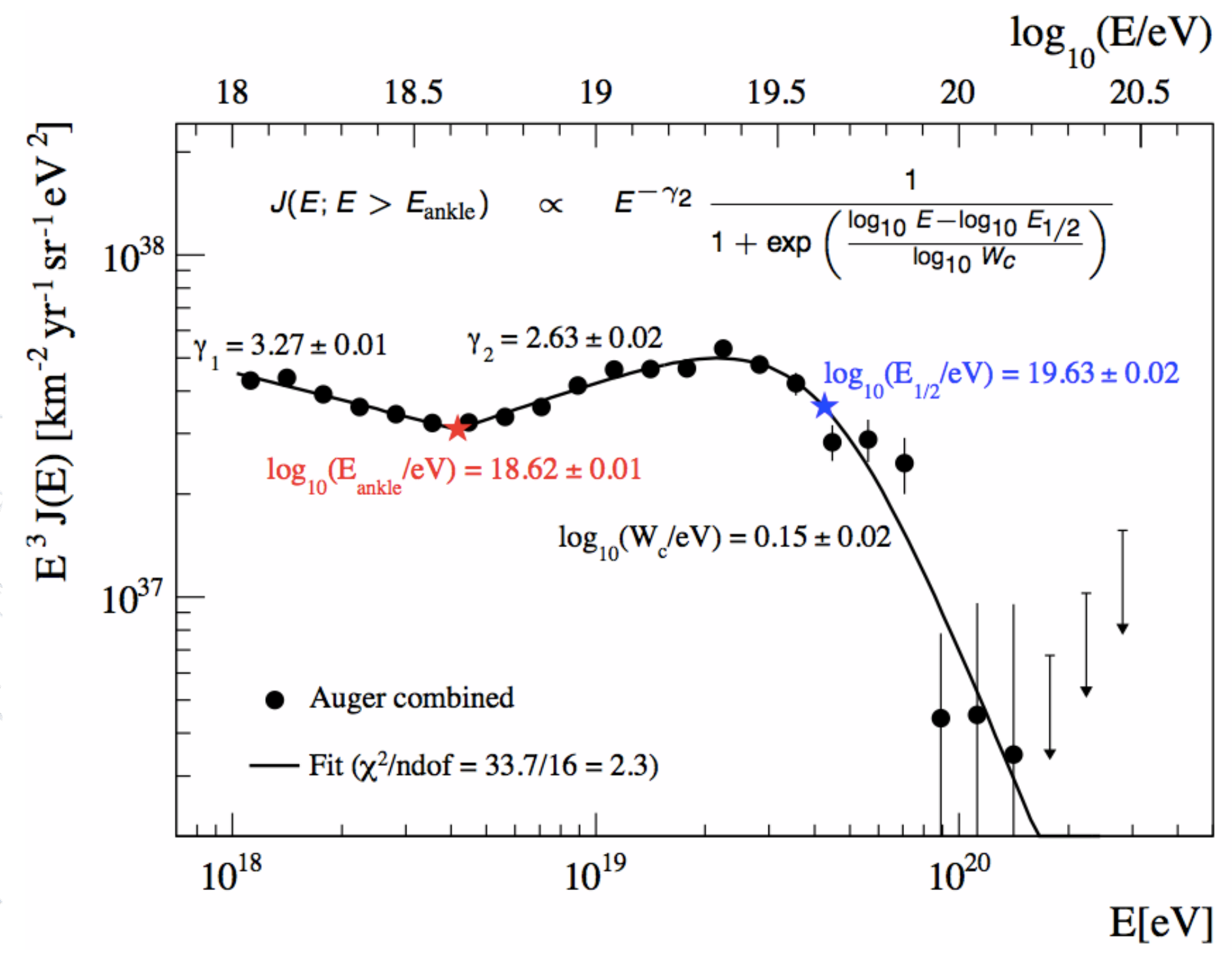}
\caption{Energy spectrum from the Auger Observatory combining hybrid and SD-only data. The error bars correspond to the statistical uncertainty. The black line is a fit with two power-laws and the smooth function written in the inset of the figure. The transition regions are indicated: the ankle (red star) and the suppression of the flux (blue star).
\label{fig:spectrum}}
\end{figure}

The features of the cosmic ray energy spectrum are sensitive to the UHECRs production mechanisms, distribution of sources and propagation effects. The Pierre Auger Observatory reported measurements of the UHECRs arrival flux at Earth, combining SD and hybrid events~\cite{spectrum}. The results are presented in figure \ref{fig:spectrum}.

The SD events used to build this plot, were collected between 1 January 2004 and 30 December 2010, with a total exposure of 20905 km$^2$ sr yr, while the hybrid events were taken between 1 November 2005 and 30 September 2010. The higher energies in the spectrum are dominated by the SD events, as it has a higher exposure. The lowest energy region benefits from the hybrid technique. In this case the exposure is calculated using the periods of telescope operation and Monte Carlo simulations to obtain the aperture. The hybrid and SD-only energy spectra are combined using a maximum likelihood fit, providing a very accurate description of the UHECR spectrum. The error bars in this figure represent the statistical uncertainty at each energy. 

The spectrum was fitted with a smooth broken power-law function that allows one to identify the main characteristics of the energy spectrum. It is possible to see that there are two clear spectral features: the ankle at $\log_{10} (E_{ankle}/{\rm eV}) = 18.62 \pm 0.01$, and a strong suppression above $\log_{10}(E/{\rm eV}) = 19.63 \pm 0.02$. The physical description of the first structure is not settled yet, but is usually accepted to be related to the transition from a galactic to an extra-galactic dominant component, leading to a change in the spectral index from $\gamma_1 = 3.27 \pm 0.01$ to $\gamma_2 = 2.63 \pm 0.02$. The second feature of the spectrum is consistent with the GZK suppression predicted by Greisen, Zatsepin and Kuz'min \cite{Greisen66,Zatsepin66}, in which the interaction of high energy cosmic rays with the CMB imposes a cutoff at the highest energies. However, since the sources that produce UHECRs are still unknown, the possibility that the sources have reached their maximum acceleration energy cannot be ruled out.

\section{Mass Composition}
The UHECRs mass composition is another key aspect to understand their origin and propagation. However, it has to be inferred through observables related to the EAS development. In Auger, one of the most sensitive variables is the depth of the shower maximum, $X_{max}$. Proton induced showers will have, on average, deeper $X_{max}$ with larger fluctuations, with respect to iron primaries. In Auger, the $X_{max}$ can be measured directly by the FD looking at the longitudinal energy deposit profile on an event-by-event basis. The statistics is however limited at the highest energies. $X_{max}$ can also be inferred from related variables at ground.

\begin{figure}[h]
\centering
\includegraphics[width=1.0\textwidth]{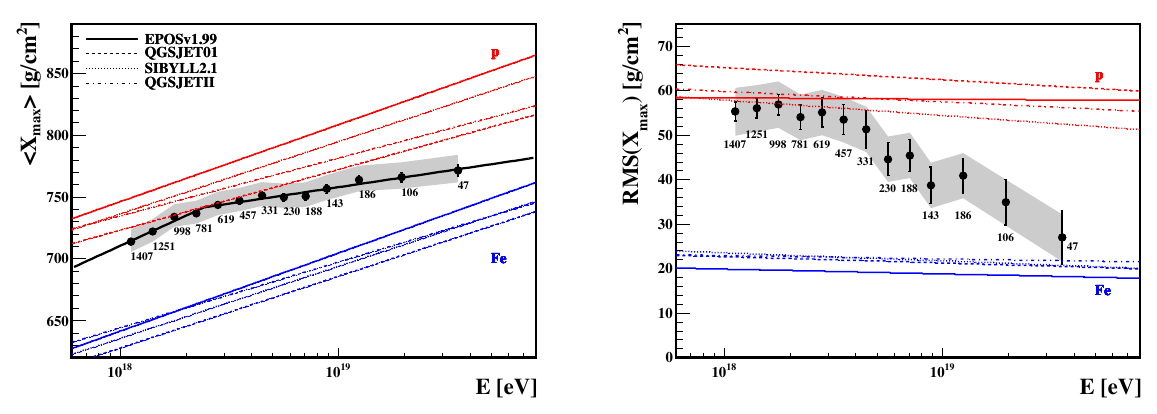}
\caption{Measurements of $\left< X_{max}\right>$ (left) and $RMS(X_{max}$ (right) as a function of the shower energy. The predictions of different high energy hadronic interaction models are shown in red (upper set of lines) for proton primaries and in blue (lower set of lines) for iron induced showers.
\label{fig:Xmax}}
\end{figure}

The average of the $X_{max}$ distribution measured by the FD, $\left< X_{max} \right>$, and the $RMS(X_{max})$~\cite{Xmax}, are shown in figure \ref{fig:Xmax} (left and right plots respectively). The depth of shower maximum depends on the hadronic interactions that rule the shower development. Therefore, the predictions of different hadronic interaction models for these observables are shown, both for proton and iron induced showers. One can observe that there is a trend towards a heavier composition at the highest energies, even though the analysis of $\left< X_{max} \right>$ and $RMS(X_{max})$ suggests a complex mass composition scenario~\cite{petrera}. In other words, a simple mixture off proton and iron does not provide a good fit for the evolution with energy of the two observables simultaneously, when using standard source models and current assumptions in the hadronic interaction models.

SD observables, such as the asymmetry of the SD signal risetime~\cite{asymm} are sensitive to primary mass composition, but on a statistical basis. Additionally, in Auger the reconstruction of the muon production depth profile can be done using the SD, on an event-by-event basis for inclined events~\cite{MPD}, with zenith angle around $60^\circ$. This is done using the shower geometry and the muons arrival times to the ground, with respect to the shower front. The depth of the maximum of this profile, $X_{max}^{\mu}$, provide another observable sensitive to the mass composition. At present, these SD observables support at the highest energies the trend measured with the FD.

The new hadronic interaction models, which have been re-tuned to describe the LHC data (EPOS-LHC and QGSJET-II-04), are now more similar to each other, but the general trend with respect to data remain unchanged. 

\section{Hadronic Interaction Properties}
Mass composition claims are intrinsically related to the description of the hadronic interactions that rule the shower development. The new data from the LHC provide an opportunity for a retuning of the parameters used in the phenomenological interaction models. Nevertheless, one must keep in mind that hadronic interactions in EAS can occur at energies up to one order of magnitude above, and in a completely different kinematic regime and collision system with respect to those provided by man-made accelerators. It is thus important to try to extract the hadronic interaction properties from the EAS data collected in Auger. Two examples will be provided: the measurement of the proton-air cross-section at $\sqrt{s} = 57$ TeV; and the measurement of the number of muons at ground. 

\subsection{Cross-Section}
\begin{figure}[h]
\centering
\includegraphics[width=1.0\textwidth]{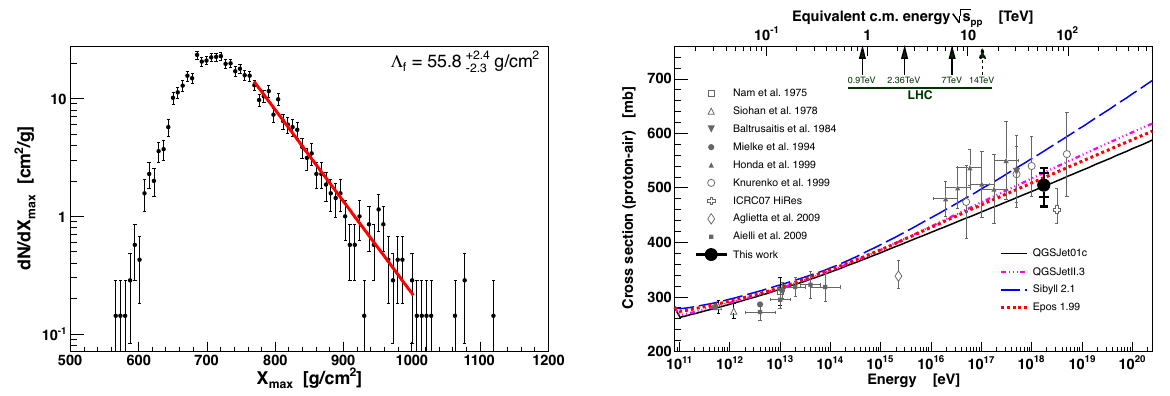}
\caption{Left: Unbinned likelihood fit of the tail of the $X_{max}$ distribution. Right: Auger measurement of the proton-air cross-section, $\sigma_{p-air}$, compared to other measurements and model predictions. The inner error bars correspond to the statistical error only while the outer bars include all systematic uncertainties for a helium fraction of $25\%$ and for the uncertainty in the fraction of photons.
\label{fig:CrossSection}}
\end{figure}

The exponential tail of the $X_{max}$ distribution at a given energy (see figure \ref{fig:CrossSection} (left) as an example), is sensitive to the first interaction point, i.e., to the primary cross-section. This distribution can be measured with the FD. Specific fiducial cuts must be applied to avoid creating a bias in the tail. 

Among the different possible hadronic primaries, protons have the smallest cross-section with air, which means that the proton component should dominate the range with the highest $X_{max}$. Therefore, the slope, obtained from an unbinned likelihood fit to the exponential tail of the $X_{max}$ distribution, (figure \ref{fig:CrossSection}, left) can be used as an estimator for the $\sigma_{p-air}$ through Monte Carlo simulations. The procedure consists of rescaling consistently the original cross-section to reproduce the value of the measurement. The main uncertainty of this analysis comes from the lack of knowledge of the mass composition at these energies. Hence, the analysis is done using events with an energy between $10^{18}$ eV and $10^{18.5}$ eV, with an average centre-of-mass energy of $57$ TeV. The $X_{max}$ measurements indicate that this region is dominated by a light nucleus composition. 

The lack of knowledge of the Helium and photon fractions are the main source of systematics on the determination of $\sigma_{p-air}$. The latter can be estimated through dedicated photon analysis, leading to an uncertainty of $10$ mb, while the He fraction is assumed to be less than $25\%$, adding an uncertainty of 30 mb. The Auger measurement of the proton-air cross section~\cite{cross-section} is of $\sigma_{p-air} = 506 \pm 22 \text{(stat)}^{+20}_{-15}\text{(sys)}$ mb, and it is compared to other measurements as well to different hadronic interaction model predictions in figure \ref{fig:CrossSection} (right).

Applying standard Glauber formalism it is possible to transform $p-air$ production cross-section into the proton-proton cross-section, and it is interesting to note that the Auger data point is within $1\sigma$ of the $\sigma_{pp}$ best extrapolation from the recent LHC data points~\cite{totem}. 

\subsection{Number of Muons at Ground}
The muon content of the shower is not only sensitive to the mass composition of the primary but also an important tool to probe the hadronic interactions that occur during the shower development, as muons are the direct decay product of mesons (mainly pions and kaons). Once muons are produced they have a large probability of reaching the ground without interacting.

\begin{figure}[h]
\centering
\includegraphics[width=0.95\textwidth]{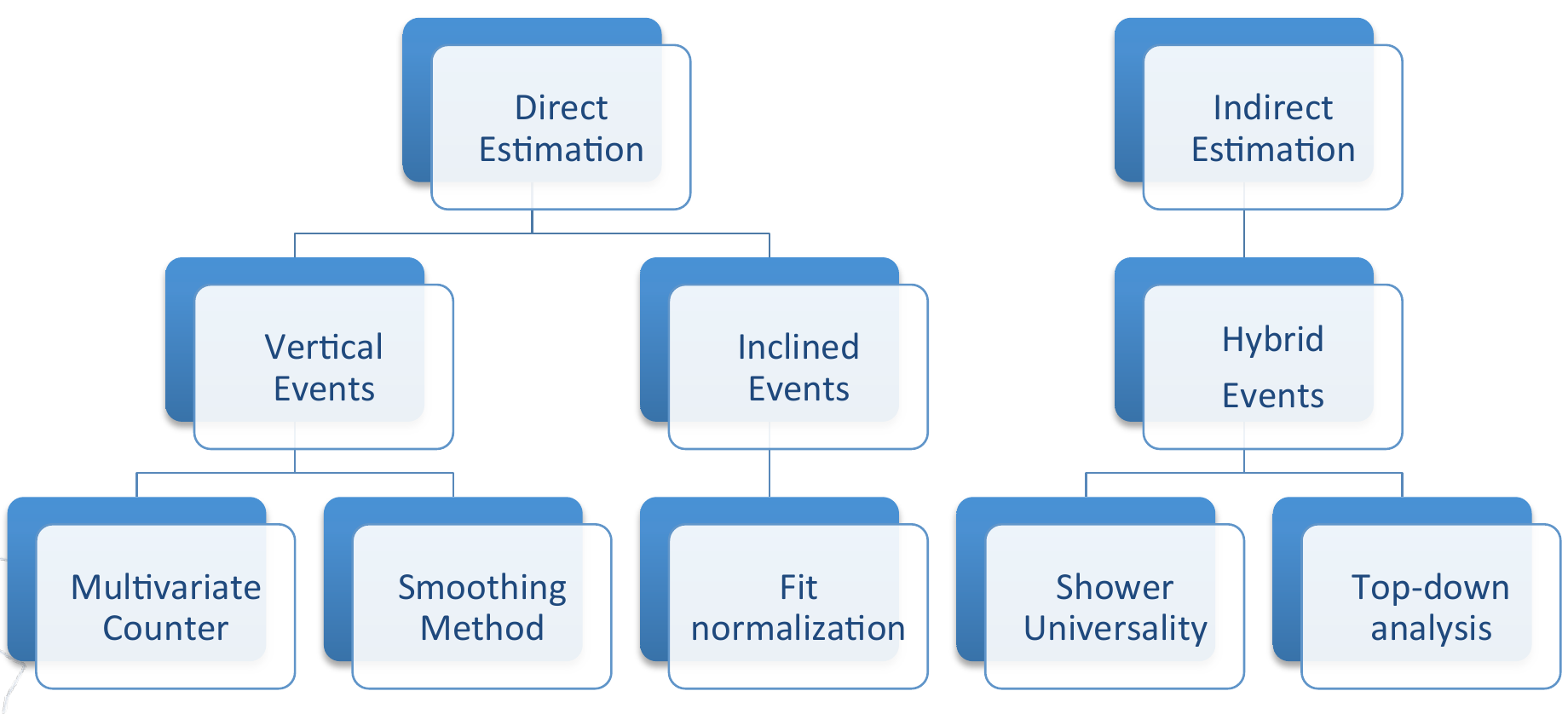}
\caption{Diagram showing the strategies applied at the Pierre Auger Observatory to measure the number of muons at the ground.
\label{fig:NmuScheme}}
\end{figure}

In Auger there are several strategies to measure the number of muons at the ground, as shown in figure~\ref{fig:NmuScheme}. Firstly one can distinguish direct and indirect measurements. The direct measurements rely on the analysis of time traces of the SD stations. For inclined showers ($\theta > 60^\circ$), the signal measured by an SD station is dominated by muons, as most of the electromagnetic component is attenuated during the shower development in the atmosphere. Hence, the number of muons can be extracted by fitting the shower footprint at the ground with simulations~\cite{inclined}. For vertical events ($\theta < 60^\circ$), the measurement can be done either by identifying muons or, by subtracting the electromagnetic component from the total signal ADC trace. Muon counting is achieved using a multivariate analysis that was trained with simulations to identify the muon peaks in the SD traces. The second approach is the so-called smoothing method which is based on the fact that the electromagnetic component has a continuous (smooth) signal in time that can be identified and subtracted using a filter algorithm. 

Indirect methods take advantage of the hybrid technique., i.e., combine FD and SD information. Two different analysis have been performed: One uses the Universality method, which is based on the fact that the muonic signal parameterised as a function of $X_{max}$ and $S(1000)$ is, according to EAS Monte Carlo simulations, independent of the primary mass composition and depends only weakly on the hadronic interaction model. The Top-down analysis, in turn, explores the information contained in high quality hybrid events. The longitudinal profiles measured by the FD are fitted with simulations for proton and iron primaries. The simulated profiles that better match the measured ones are then used to predict the signal at ground level and afterwards compared with the SD data. 

\begin{figure}[h]
\centering
\includegraphics[width=0.95\textwidth]{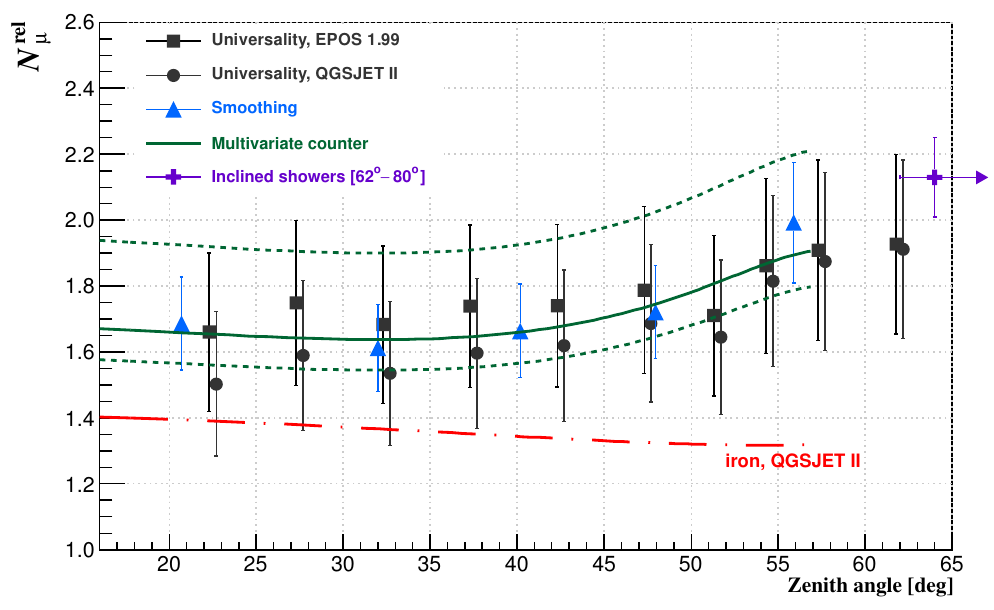}
\caption{Number of muons estimated at 1000 m from the shower core relative to the predictions of simulations using QGSJet-II.03 with proton primaries at $E = 10^{19}$ eV. The results are displayed as a function of the zenith angle.
\label{fig:Nmu}}
\end{figure}

The number of muons, estimated at 1000 m from the shower core, relative to the predictions of simulations using QGSJet-II.03~\cite{QGSII} with proton primaries, $N_{\mu}^{\rm rel}$, is shown in figure~\ref{fig:Nmu} for the different analysis methods listed above\footnote{except for the top-down analysis whose comparison with all the other methods results cannot be done directly. Nevertheless, also in this analysis there can be seen a deficit of the number of muons in simulation with respect to data.}~\cite{Nmu}. All the methods present compatible results, within uncertainties. The number of muons predicted by the models shows a deficit with respect to data, and the deficit increases with zenith angle. This deficit is present even choosing iron primaries and the hadronic interaction model characterised by having more muons (EPOS1.99~\cite{EPOS}). The main factors that affect this discrepancy between data and Monte Carlo simulations are the uncertainty on the energy scale (currently around $22\%$), the unknown mass composition and uncertainties on the hadronic models (for instance, potential problems on the muon attenuation). However, none of these provides an easy solution by itself. 

\section{Searches for Photons and Neutrinos}
The Pierre Auger Observatory can in addition be used to search for neutral primaries, in particular photons and neutrinos. Primary photons are expected in non-acceleration models of UHECRs origin (top-down scenarios), where, for instance, topological defects or heavy dark matter decays could occur near Earth, producing high energy cosmic rays. 

\begin{figure}[h]
\centering
\includegraphics[width=0.95\textwidth]{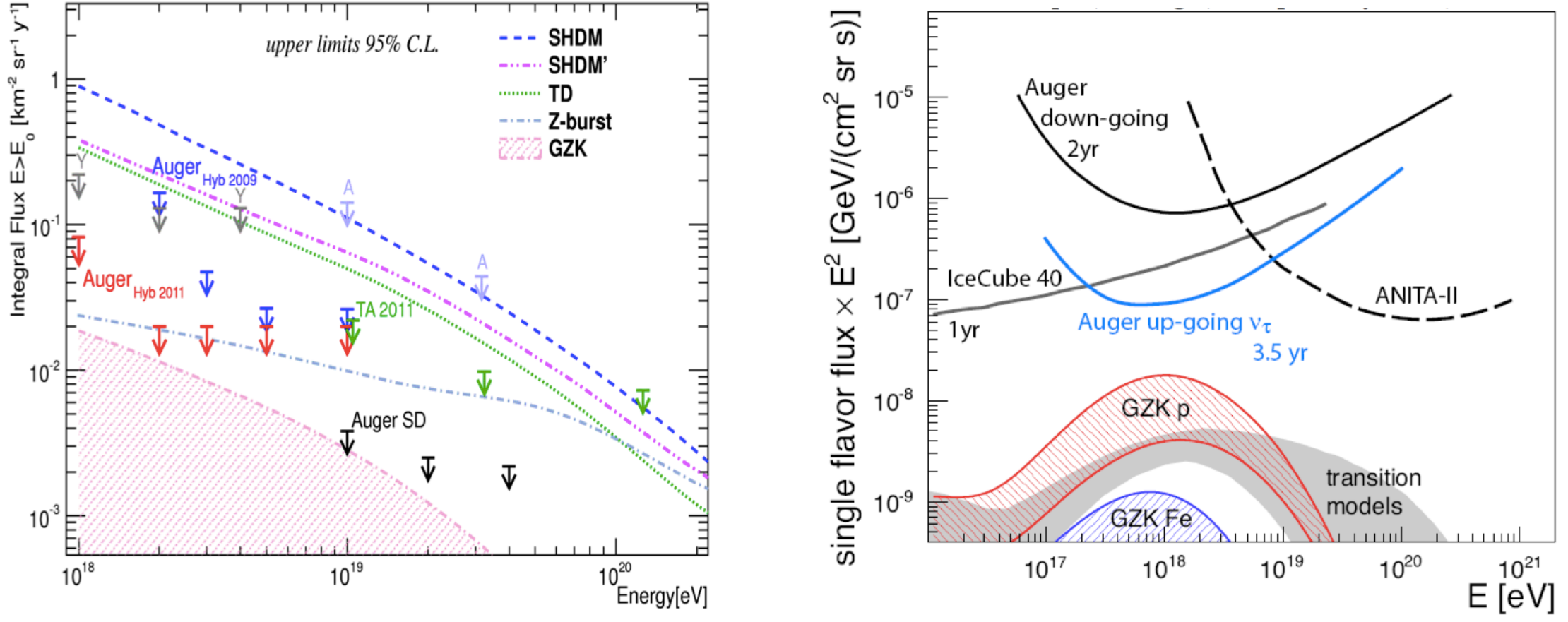}
\caption{Left: Upper limits on the photon flux derived using hybrid analysis (red arrows) and SD data (black arrows). The shaded regions correspond to GZK predictions for primary photons while the lines are for exotic models. Right: Upper limits on the diffuse high-energy neutrino flux. The shaded areas are predictions for GZK neutrinos for different primaries. The reach of other experiments is also shown for comparison~\cite{unger}.
\label{fig:photonNeutrino}}
\end{figure}

At a given energy, photon initiated showers are expected to develop slower than proton induced showers, due to smaller secondary multiplicities and to the suppression of the cross-section due to the LPM effect. Therefore, one can search for photon events looking for events with an unexpectedly large $X_{max}$ measured by the FD.  Shower parameters at the ground, such as the radius of curvature and the signal rise time can also be used as discriminators for photons. They should be larger and slower respectively for photons than for proton induced showers. The upper limits on the photon fluxes are shown in figure \ref{fig:photonNeutrino} (left). This result significantly disfavours the top-down scenarios~\cite{photons}. 

Neutrinos can be detected using the SD by searching for very inclined showers (near the horizon). Due to the very low interaction cross-section of a neutrino, it can traverse a large amount of atmosphere and produce a shower near the array. This shower is relatively \emph{young}, thus having both muonic and electromagnetic components. In contrast, a hadronic induced shower would start early in the atmosphere, and would appear in the array as an \emph{old} shower. Due to the high zenith angle, the electromagnetic component would be severely attenuated and the shower would only have muonic component. The analysis is based on the width of the time distribution signals recorded by the SD stations. \emph{Old} showers have a narrow signal when compared to \emph{young} showers. Another way to detect neutrinos consists in looking for $\nu_{\tau}$ that traverse the earth, starting an up-going shower of particles near the array (earth skimming events). Up to now, no candidate neutrino events have been observed~\cite{neutrinos}, resulting in upper limits on diffuse neutrino fluxes given in figure \ref{fig:photonNeutrino} (right). 

Expected secondaries of the GZK effect include both photons and neutrinos, as shown on the shaded bands in figure \ref{fig:photonNeutrino}. These very low fluxes could be reached in the next years of Auger operation, providing an additional signature for the presence of the GZK cutoff. 

\section{Conclusions}
The Pierre Auger Observatory is the world largest observatory to study UHECRs and is smoothly accumulating data at the highest energies. It has measured the end of the cosmic ray energy spectrum with unprecedented statistics, confirming the presence of an ankle and of a suppression at the highest energies.

The mass composition analyses indicate a trend towards heavy composition as the energy increases. However, any interpretation depends on the understanding of high energy hadronic interactions, and the measurement of the number of muons at $E=10$ EeV ($\sqrt{s} \sim 130$ TeV) indicate that the models may be incomplete as their predictions present a deficit of muons.

Auger has measured the proton-air cross-section at $\sqrt{s} = 57$ TeV to be $\sigma_{p-air} = 506 \pm 22 \text{(stat)}^{+20}_{-15}\text{(sys)}$ mb. It has also put upper limits on the flux of photons and neutrinos. The current limits on primary photons severely disfavour top-down scenarios for the origin of UHECRs.

The Pierre Auger Observatory offers a unique opportunity, not only to study the origin and composition of the highest energetic particles known, but also to study particle physics at energies well above current human-made accelerators. 

\section*{Acknowledgements}
I would like to thank to L. Apolin\'{a}rio, M. C. Esp\'{i}rito Santo and M. Pimenta for carefully reading the manuscript. This work is funded by FCT, Funda\c{c}\~ao para a Ci\^encia e Tecnologia (SFRH/BPD/73270/2010).

\bibliographystyle{woc}
\bibliography{Bib-Auger}

\end{document}